\documentclass[11pt,legal]{article}
\usepackage[utf8]{inputenc}
\usepackage{theorem,ifthen,algorithm,algorithmic}
\usepackage{amssymb,amsfonts,amsmath,latexsym,dsfont}
\usepackage{tikz,pgflibraryplotmarks}
\usepackage{fullpage}
\usepackage{graphicx}
\usepackage{mathrsfs}
\usepackage{hyperref}
\usepackage[font=small,labelfont=bf]{caption}
\usepackage{multirow}
\usepackage{color, colortbl}

\definecolor{Gray}{gray}{0.9}

\newcommand{\calR}{{\mathcal{R}}}
\newcommand{\calT}{{\mathcal{T}}}
\newcommand{\calM}{{\mathcal{M}}}

\newcommand{\R}{\ensuremath{\mathds{R}}}

\graphicspath{{images/}{./}}

\title{Comparison of atlas-based and neural-network-based semantic segmentation for DENSE MRI images}

\author{Elle Buser\thanks{Department of Mathematics and Statistics, University of Wyoming, Laramie, WY, USA},\; Emma Hart\thanks{Department of Mathematics, Colgate University, Hamilton, NY, USA},\; Ben Huenemann\thanks{Department of Mathematics, University of Utah, Salt Lake City, UT, USA}\\ Project Advisors: Lars Ruthotto\thanks{Department of Mathematics and Department of Computer Science, Emory University, Atlanta, GA USA}, Justin Smith\thanks{Barack Obama Magnet Elementary, Atlanta, GA, USA}}

\begin{document}
\maketitle

\begin{abstract}
    Two segmentation methods, one atlas-based and one neural-network-based, were compared to see how well they can each automatically segment the brain stem and cerebellum in Displacement Encoding with Stimulated Echoes Magnetic Resonance Imaging (DENSE-MRI) data. The segmentation is a pre-requisite for estimating the average displacements in these regions, which have recently been proposed as biomarkers in the diagnosis of Chiari Malformation type I (CMI).
    In numerical experiments, the segmentations of both methods were similar to manual segmentations provided by trained experts.
    It was found that, overall, the neural-network-based method alone produced more accurate segmentations than the atlas-based method did alone, but that a combination of the two methods -- in which the atlas-based method is used for the segmentation of the brain stem and the neural-network is used for the segmentation of the cerebellum -- may be the most successful.
\end{abstract}

\section{Introduction and Problem Outline}

Semantic segmentation is the process of identifying specific regions of an image by labeling each pixel as being part of a class. This process is has a wide range of applications in image analysis, from the development of self-driving vehicles -- that use computer vision to understand where roads, pedestrians, and other vehicles are -- to organizational image search tools -- that allow users to automatically sort their pictures by content. The enormous amounts of image data generated each day by people around the world provide the field of semantic-segmentation with an ever-growing domain of both data and possible application.

One application of semantic segmentation in medical imaging is in the diagnosis of Chiari Malformation type I (CMI), a condition affecting the skull and brain that is estimated to affect slightly fewer than 1 in 1,000 people \cite{AANS}. Although most of these cases are asymptomatic, some can lead to extreme pain and require immediate surgery. In such cases CMI can be accompanied by severe head and neck pain, headaches, dizziness, and impaired vision \cite{BologneseEtAl2019}.

Although CMI is typically diagnosed when an individual's cerebellar tonsils (TP) extend more than 5mm below the foramen magnum, many experts recommend rejecting this criterion due to a weak correlation between CMI and TP \cite{BologneseEtAl2019}. A recent study also found that there is a significant increase in neural tissue movement in CMI patients compared to controls, especially in the brain stem and cerebellum \cite{NwotchouangEtAl2020}. This was found using a dynamic MRI technique called Displacement Encoding with Stimulated Echoes (DENSE): a type of MR imaging that records information about movement in the brain and that is sensitive enough to capture micrometer scale deformations that occur with each heartbeat. Measurements related to the displacement in the brain stem and cerebellum could perhaps become part of a new, more reliable basis for the diagnosis of CMI.

\begin{figure}[t]
    \centering
    \includegraphics[width=.8\textwidth]{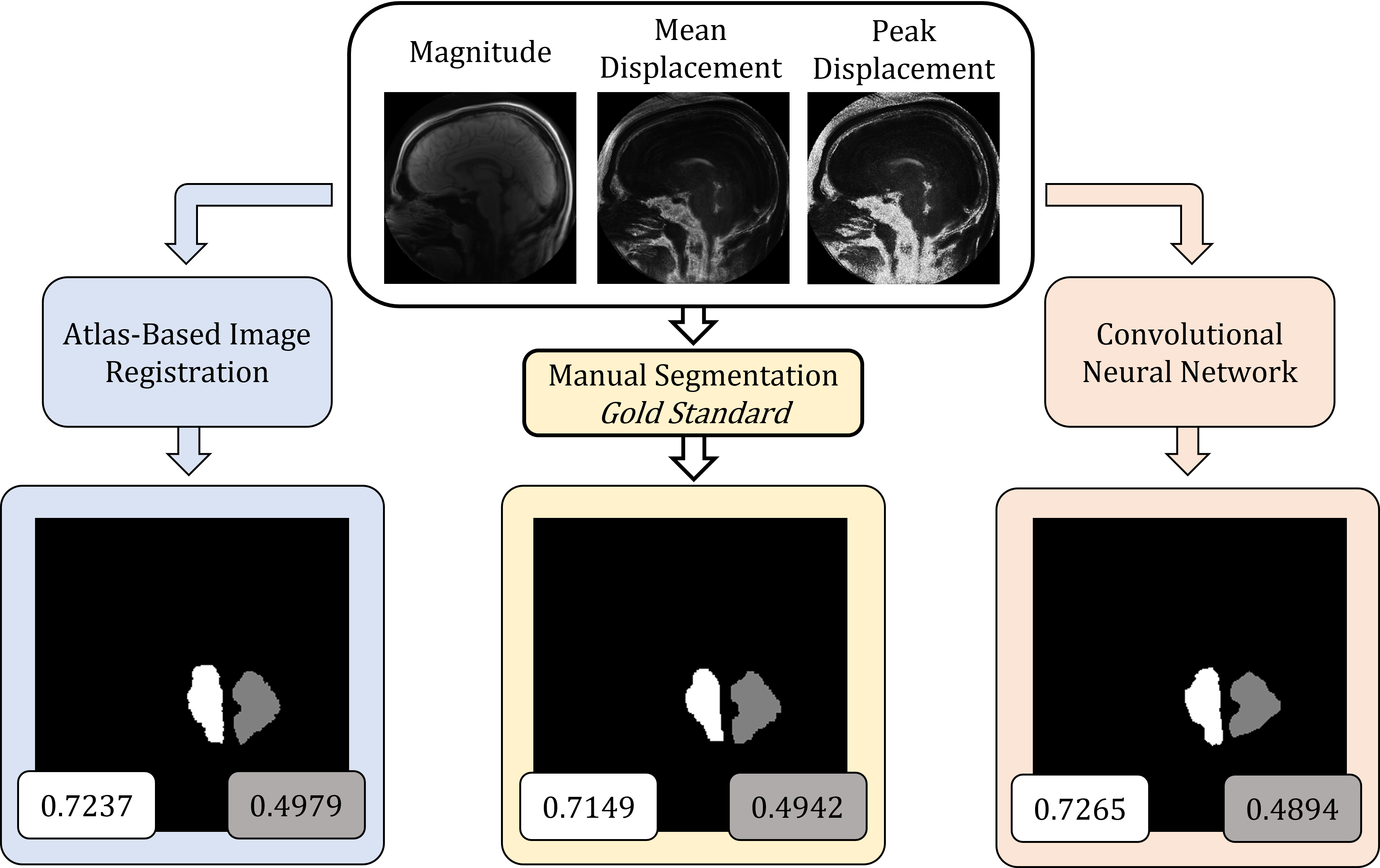}
    \caption{An overview of the setup of this study. The current method of diagnosing Chiari from brain displacement data (center gold path) involves a radiologist manually drawing a mask that identifies the brain stem and cerebellum, and computing the spatial average and temporal peak displacement on those regions as a biomarker for CMI diagnosis. The brain stem peak-displacement biomarker is shown for each method in the white box on the left, while the cerebellum's is shown in the grey on the right. An atlas-based algorithm (left blue path) and a neural-network-based approach (right red path) were developed in this study to segment the cerebellum and brain stem given DENSE MR imaging and produce biomarkers automatically. Results were compared between the two methods and to the results produced from the manual process, that are treated as truth.}
    \label{fig:paper-overview}
\end{figure}

The purpose of this study is to create image segmentation algorithms that automatically identify these regions and conduct further analysis with DENSE MRI data. One of the most limiting steps in diagnosis based on this new criterion is the accurate segmentation of the brain stem and cerebellum in the DENSE-MRI data. Currently, the brain regions are labeled manually, which is a time consuming and, thus, expensive process \cite{NwotchouangEtAl2020}. This study compares two distinct methods to automate the segmentation of the brain stem and cerebellum: an atlas-based image registration approach, and a convolutional-neural-network-based approach (paper overview shown in \textbf{Figure \ref{fig:paper-overview}}).

The data set that is used to automate this process includes DENSE MR images and masks identifying the cerebellum and brain stem for patients from Emory University Hospital's Department of Radiology and Imaging Science.
The data set has been split into training data, which is used to create of the neural-network-based and atlas-based segmentation methods, and testing data, which is used to gauge the generalization of the methods to unseen data.

In investigating atlas-based image registration, our study builds on methods from a MATLAB toolbox called FAIR \cite{Modersitzki2009}. Image registration from this toolbox relies on the idea of a template image and a reference image. In the case of segmenting MR images, the reference is the magnitude image (described further in \textbf{Section \ref{data subsection}}) that needs segmentation and the template is some magnitude image that has already been segmented. The main idea of this approach is to find a transformation that approximately aligns the template image to the reference.

Another approach is found in convolutional neural networks (CNNs), a category of machine learning. Using MR images from the data set as input, a model is trained in a supervised fashion to output a segmentation with three classes: brain stem, cerebellum, and background. In this study, a CNN made for biomedical imaging called U-Net \cite{RonnebergerEtAl2015} is implemented the python-based machine learning library PyTorch \cite{paszke2017automatic}. This network, combined with an optimizer and loss function, loops over a section of the data set, called the training set, to produce an output mask by classifying each pixel as background, brain stem, or cerebellum. This output is then compared to the corresponding known mask. With each iteration, the CNN learns from this comparison and updates its parameters to produce more accurate results. The model can then be tested on images in the data set, not used in training, as a means to see how well it segments new MR images.

Overall, as a single method of semantic segmentation, the neural-network based approach outperformed the atlas-based approach.  However, both methods worked better for different regions; while the neural-network based method was able to produce a more accurate biomarker for the cerebellum in a majority of cases, the atlas-based method was able to produce a more accurate biomarker for the brain stem in a majority of cases. Together, for our testing set, the methods were able to produce segmentations of the brain stem and cerebellum that were, on average, 86\% similar to radiologist drawn segmentations (measured with Dice similarity index, described in \textbf{Section \ref{dice subsection}}) and that were able to predict the manually produced biomarkers for the brain stem and cerebellum with an average 4\% relative error each.

\section{Problem Description}
This section begins with a detailed description of the dataset used, then an introduction to the measures used to evaluate both segmentation and diagnosis. This is followed, in the next sections, first by an explanation of the atlas-based segmentation method used, then by an explanation of the convolutional neural-network-based approach.

\subsection{Data and Pre-Processing}
\label{data subsection}
This study uses 256x256 pixel DENSE MR image data and associated masks for 63 patients.
We randomly split the data into a training set, containing 51 examples used in the creation of the neural-network-based and atlas-based methods, and and a test set, consisting of 12 examples which we use to gauge the generalization of the methods. As is typical in MR imaging, the DENSE MRI data is stored as a complex-valued array.

\begin{figure}[t]
    \centering
    \includegraphics[width=\textwidth]{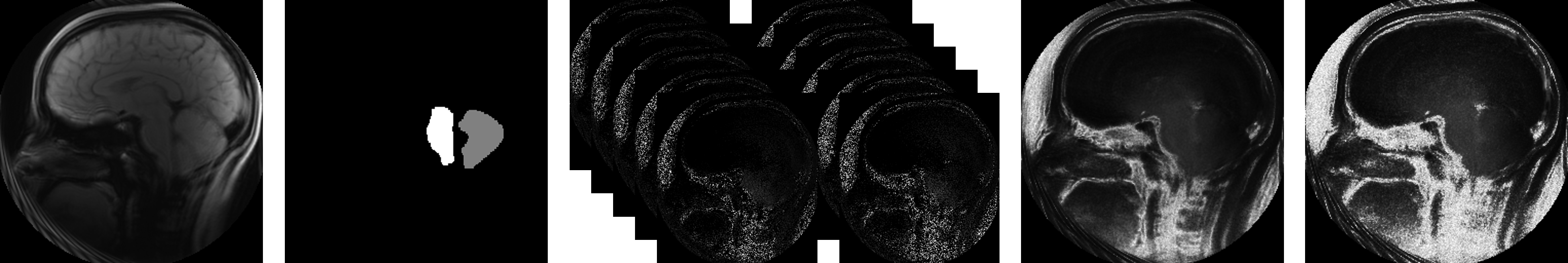}
    \caption{From left to right, an example of one patient's (1) magnitude MRI image; (2) mask; (3) DENSE images representing one cardiac gate each; (4) temporal mean DENSE image; and (5) temporal peak DENSE image.}
    \label{fig:data-example}
\end{figure}

The images typically visualized in MRI are based on magnitude, where the grey value of each pixel is created from the magnitude of the complex value in that pixel’s position (shown in \textbf{Figure \ref{fig:data-example}}(1)). The color of each of the pixels in these magnitude images is controlled by the chemical makeup of the material that exists in that position in the brain, and therefore represent the brain's anatomy. 

\begin{figure}[t]
    \centering
    \includegraphics[width=.6\textwidth]{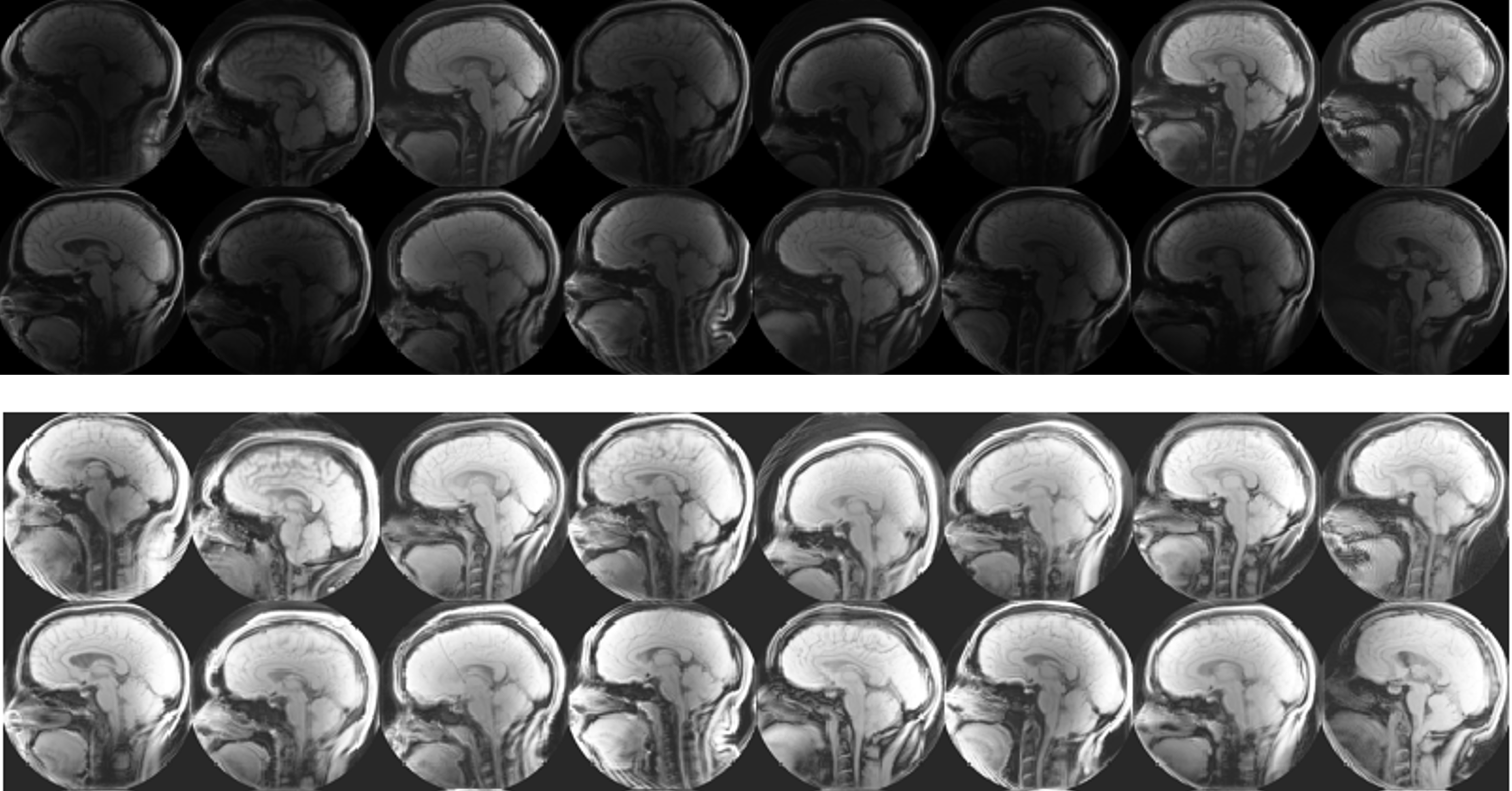}
    \caption{A sample of sixteen magnitude images, original (top) and normalized using histogram normalization (bottom).}
    \label{fig:normal}
\end{figure}

Since the brightness and contrast levels varied across the magnitude images in a way that made segmentation more difficult, we pre-processed the data using histogram normalization. To this end, we used the MATLAB function histeq, which approximately linerarizes the cumulative distribution function and helps create a more uniform set of images. \textbf{Figure \ref{fig:normal}} compares a sample of original and normalized images. For more explanation of this process, see \cite{Coste}.

DENSE imaging encodes displacement information in the phase of each complex number corresponding to the intensity of movement associated with each pixel position. For each subject many DENSE MRI scans were created corresponding to gates that occur over the patient's cardiac cycle (shown in \textbf{Figure \ref{fig:data-example}}(3)). The number of DENSE images varies per patient based on their heartbeat. The visual interpretation of these phase images is perhaps less intuitive than the magnitude images, mapping movement rather than anatomy.

Generally, areas of large movement in brain tissue appear lighter than those of small movement, but in areas of especially high movement -- especially outside the brain where cerebrosprinal fluid (CSF) moves much more quickly than the micrometer deformations in the brain tissue -- phase wrapping occurs. Because there is no distinction between the gray scale color of a pixel stored with a phase $\phi$ and $\phi + 2\pi$, for example, they appear the same. These regions with high movement appear as random noise because the movement is beyond the smaller scale for which the phase encoding was calibrated. For more on phase wrapping, see \cite{phasewrap}.

In this study, we follow  \cite{NwotchouangEtAl2020} and remove the temporal dimension of the DENSE MRI data by computing a mean DENSE image, showing the mean value per pixel over time (\textbf{Figure \ref{fig:data-example}}(4)), and a peak DENSE image, showing the maximum value per pixel over time (\textbf{Figure \ref{fig:data-example}}(5)), for each subject.
This representation of the DENSE MRI data over time also helps reduce the regions of random noise and summarize over the patient's whole cardiac cycle. For more discussion of this choice and of other ways to represent this data, see \cite{NwotchouangEtAl2020}.

The masks associated with the magnitude images classify each pixel as being part of the background, the cerebellum, or the brain stem, and were drawn by researchers in Emory's Department of Radiology and Imaging Science (shown in \textbf{Figure \ref{fig:data-example}}(2)) as part of the study~\cite{NwotchouangEtAl2020}.

\subsection{Notation}
\label{Notation}
The neural-network-based and atlas-based approaches use image data in different ways. The first keeps the images as MRI discrete data while the latter uses these data to build a continuous model through interpolation. 

In the neural-network-based method, the input, $I \in \R^{256 \times 256 \times 3}$, is made of 3 channels: the normalized magnitude, peak DENSE, and mean DENSE images. These are each 256x256 matrices with entries corresponding to the gray values.

The atlas-based approach treats an image as an interpolated function, ${\cal I} : \Omega \to \R$ where the domain $\Omega \subseteq \R^2$ corresponds to coordinates of pixels in the image grid. The multilevel processes, described in \textbf{Section \ref{Registration}}, rely on this continuous model of our data. The range of this function, called on a specific 256x256 grid, creates the discrete 256x256 pixel images. To distinguish these understandings, calligraphic letters have been used to denote continuous functions while regular script denotes discrete. This notation, and the notation of other equations related to the atlas-based method, are formatted in the style of FAIR \cite{Modersitzki2009}. This MATLAB toolbox and its associated textbook~\cite{Modersitzki2009} can be referred to for a more complete discussion of image registration. 

These two understandings of images lend also themselves to two conceptions of the outputted masks. In the neural-network-based method, the output mask is a discrete 256x256 matrix which predicts a class $k \in \{0,1,2\}$ for each pixel where 0 corresponds to background, 1 to cerebellum, and 2 to brain stem. $M$ symbolizes the known target masks while $M_{p}$ corresponds to the model predicted mask.

Analogously, in the continuous function sense, let $C$, $B$, and $T$ be the sets representing the cerebellum, brain stem, and the total union between them respectively. Then let $\chi_C, \chi_B : \Omega \to \{0, 1\}$ be the characteristic functions for the cerebellum and brain stem respectively. Masks that display these regions will hence be defined as $\calM : \Omega \to \{0, 1, 2\}$ where
\begin{equation}
    \calM(x) = \chi_C(x) + 2\chi_B(x).
\end{equation}
In addition to this, we define the total (T) mask characteristic function to be $\chi_T : \Omega \to \{0, 1\}$ where $\chi_T = \chi_{C \cup B}$.

Using this notation, the template mask will be referred to as $\calM_\calT$ and the reference mask as $\calM_\calR$.

\subsection{Evaluation Measures}
Our ultimate goal is to automatically segment the brain stem and cerebellum so that we could produce peak-displacement biomarkers for those regions that match radiologist produced ones. The cerebellum's average peak-displacement, for example, is calculated by averaging the peak-displacement of each pixel in the DENSE MRI that is classified as being part of the cerebellum (and likewise for the brain stem). In addition to producing accurate biomarkers we also quantify the accuracy of the segmentations to ensure the plausibility of the biomarker.

Radiologist drawn masks that identify what displacement data to average to obtain the biomarkers are treated as the gold standard. It may be that these algorithmic methods can produce better, more consistent masks, but that lies outside the scope of this study. To balance the dual concerns of producing masks and biomarkers that match manually produced ones, Dice similarity indices and biomarkers were considered at each step.

\subsubsection{Dice Index}
\label{dice subsection}
To evaluate the success of both the neural-network-based and atlas-based segmentation, we use the Dice similarity index. Given two subsets of $A,B \subset \Omega$, the Dice index is computed as
\begin{equation}\label{eq:Dice}
    D_{\rm{Dice}}[A, B]
    = \rm{Dice}[\chi_{A}, \chi_{B}]
    = 2 \int_{\Omega} \frac{\chi_{A}(x) \chi_{B}(x)}{\chi_{A}(x)+\chi_{B}(x)} dx,
\end{equation}
where $\chi_{A}(x), \chi_{B}(x): \Omega \to \{0,1\}$ represent the characteristic functions for the sets $A$ and $B$, respectively. This results in a measure that ranges from 0, in the case of two disjoint regions, to 1, in the case of identical regions \cite{Monteux2019}.

We use this measure primarily to evaluate mask similarity and determine if algorithmically predicted masks reasonably match radiologist drawn masks. Our masks define three sets of special interest: the set of pixels identified as the brain stem, the set identified as the cerebellum, and the set identified as background. Computing the Dice similarity for the cerebellum and brain stem, especially, between a true mask and predicted mask for the same brain quantifies how well the segmentation method is working; in both the neural-network-based and atlas-based approaches, algorithms and parameters were chosen to maximize this number.

\subsubsection{Biomarkers} Recently, the spatial mean and temporal maximum of the magnitude displacement measured in DENSE-MRI over the cerebellum and brain stem have shown promise as a biomarker for diagnosing CMI~\cite{NwotchouangEtAl2020}. For example, once a segmentation is identified for the brain stem, the value of each pixel classified as a part of this region from the peak-displacement DENSE image data is averaged (or analogously, for the cerebellum). Though the world ``biomarker" applies broadly, and attributes including age, weight, and sex could each be considered biomarkers that are relevant to CMI, when we refer to biomarkers throughout this paper, we are referring to the displacement averages across the brain stem and cerebellum described above. Building on the results of \cite{NwotchouangEtAl2020}, the biomarkers produced from manual segmentation will be treated as the truth, and our aim is to match these with our automated approaches.

\section{Atlas-Based Segmentation}

This section includes discussion of the way a template image is chosen, a way of averaging the segmentation results, the parameters and functions used in image registration (associated with the toolbox FAIR \cite{Modersitzki2009}), and examples of the method in full.

\subsection{Overview of the Registration Pipeline}
\begin{figure}[t]
    \centering
    \includegraphics[width=.7\textwidth]{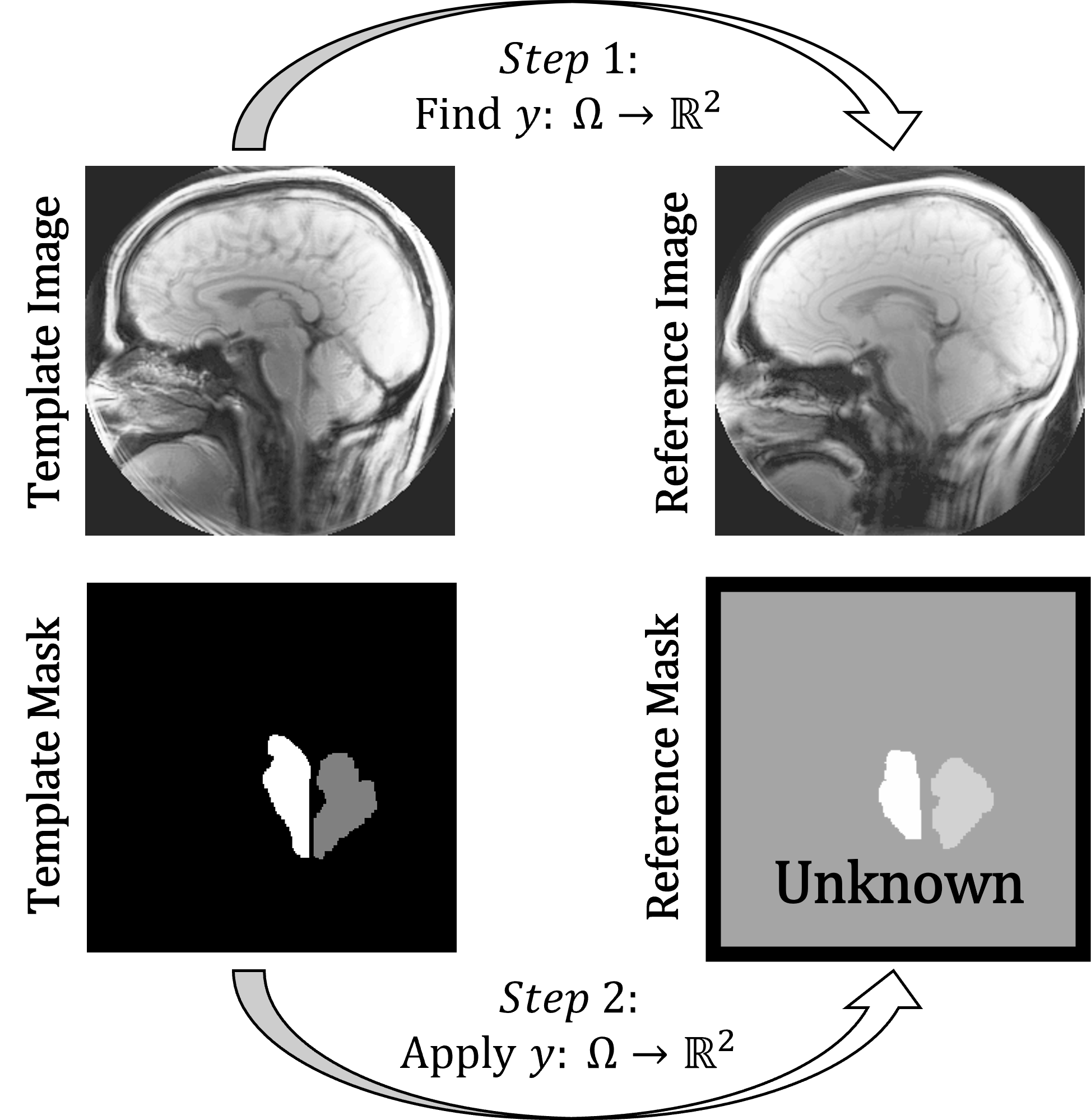}
    \caption{A general overview of the registration process, that involves finding a transformation between a template and reference and then applying that transformation to the template mask to predict a reference mask.}
    \label{fig:atlas-example}
\end{figure}
Given a new image without a mask, our method compares the new image to a set of 51 labeled images in order to predict a mask for the new image. To allow for a pixel-to-pixel comparison, we solve a set of registration problems. Following the language used in FAIR, the new image is called the reference, while the known labeled images are templates. 

Here, we discuss how templates are chosen to be compared to new references along with an averaging method to reduce outlying results, and in the next section we will describe the registration performed in each of these comparisons.

\subsubsection{Choosing a Template Image}
\label{Template}

Perhaps the most critical factor in the success of the atlas-based segmentation is picking an appropriate template image for a new reference; when the brains in a template and reference image are already similar -- in their shape and in how they are positioned -- the registration is best set up for success. However, choosing the most similar template image is not a trivial task. While the Dice index between the template and reference masks could provide a useful measure, as it captures the alignment we are interested in, the reference mask is what we aim to produce so this index cannot be calculated before registration. For this reason, we had to rely on calculating some type of similarity metric between the magnitude images. There is a wide range of options for this that value different attributes. Three of these options that we explored are the Normalized Gradient Field (NGF), Sum of Squares Distance (SSD), and weighted Sum of Squares Distance (wSSD).

A NGF method works by aligning edges identified in the magnitude images by the gradient -- relying more on the shape and location of the brains (for more on NGFs, see  \cite[Sec. 7.4]{Modersitzki2009}). This showed a slight correspondence, but was much less successful than using SSD (this metric is described in more detail in \textbf{Section \ref{Registration}}). Using a wSSD metric allowed the program to focus on the section of the brain with the brain stem and cerebellum. While this can be effective, the main flaw was that the skull shape was an important factor that was being ignored. The final similarity metric that we decided on was regular SSD because it seemed to be the overall best and most adaptable.

\begin{figure}[t]
    \centering
    \begin{tabular}{c}
         \textbf{Dice Mask Similarity versus SSD} \\
         \includegraphics[width=0.9\textwidth]{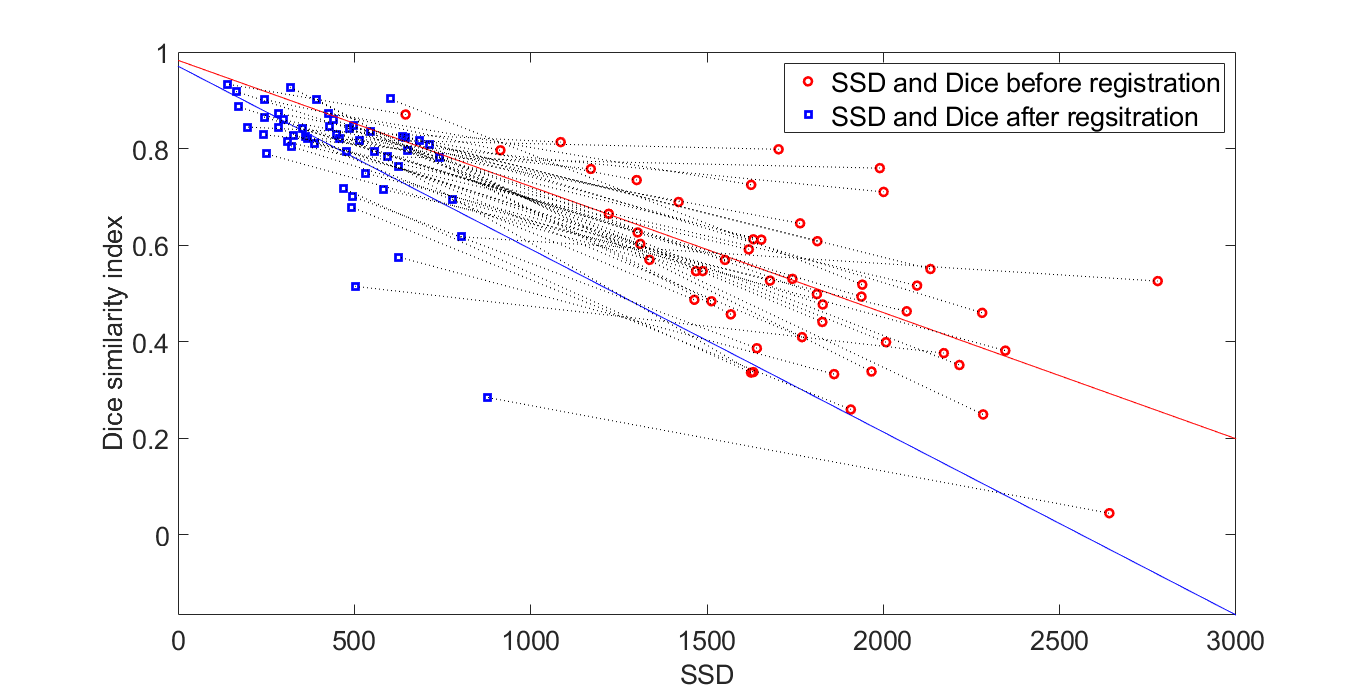}
    \end{tabular}
    \caption{Plot of the Dice similarity index and SSD between an example reference image and all template images in our training set. The red circles represent the measures before the registration is performed (with a corresponding solid red least squares line), while the blue squares represent them after it is performed (with a corresponding solid red least squares line). Black dotted lines connect points representing the same templates before and after registration.}
    \label{fig:ssd_similarity}
\end{figure}

In \textbf{Figure \ref{fig:ssd_similarity}}, 50 image registrations were run with one example patient used as reference and each of our other 50 training images used to see how well the SSD works as a similarity measure for the magnitude images, with the SSD measure plotted against the ideal, mask-similarity Dice measure. Note that whereas a higher Dice index represents higher similarity, the opposite is true for the SSD measure; if two exactly similar images were compared, the SSD would measure zero, while the Dice index would measure one. Both before and after registration, the SSD and Dice index are, to an extent, negatively correlated, which helps justify our use of it as a way to chose initially similar templates for our reference. Though it is by no means a perfect measure, for our case, it provides an intuitive and computationally efficient measure that helps improve registration results.

\subsubsection{Averaging the Registrations}
\label{Averaging}

The significance of edge cases in the registration creates a need for less dependence on any individual template image. To accomplish this, the registration was run on multiple template images and the resulting masks were averaged together. Then the resulting average mask could be converted to a binary image and used as a more definitive prediction of the segmentation.

To be more specific, the variable $n$ was used to correspond to the number of patients that the registration program is run on. Using the same reasoning behind choosing a template as \textbf{Section \ref{Template}}, the SSD is calculated for each template and the lowest $n$ are selected. These images are then summed together and divided by $n$ to get the average mask. Finally, a threshold is used to convert this average transformed mask into a binary mask again. For discussion of how the values of $n$ and threshold were chosen, see the beginning of \textbf{Section \ref{AtlasResults}}.

\subsection{Registration Algorithm}
\label{Registration}

The goal of the image registration is to find some transformation $y : \Omega \to \R^2$ such that
\begin{equation}
    \calT[y](x) \approx \calR(x), \quad \text{ for all } \quad x \in \Omega.
\end{equation}
Note that $\calT[y](x) = \calT(y(x))$ is written this way to emphasize that the function $y$ is an input as well. \textbf{Section \ref{Notation}} provides insight into other notational choices.

We find that function $y$ by minimizing the nonconvex  objective functional
\begin{equation}
    {\cal{J}}_{\rm{atlas}}[y] = D_{\rm{SSD}}[\calT[y], \calR]+ \alpha {\cal S}[y],
    \label{joint_objective}
\end{equation}
where $D_{\rm{SSD}}$ is the sum of squares distance measure, $\alpha$ is the regularization parameter, and $\cal S$ is the regularizing functional. These parameters along with some fundamental design decisions are described in more detail below: \\

\textbf{Distance}: When performing the image registration, the algorithm tries to minimize the distance measure. There are many options for how to pick this distance measure, but the perhaps simplest and most common is the Sum of Squares Distance (SSD) given by
\begin{equation}
    D_{\rm{SSD}}[\calT[y], \calR] = \frac{1}{2} \int_{\Omega} (\calT[y](x) - \calR(x))^2 dx.
\end{equation}
The SSD measure penalizes different brightness levels. As a result, altering the contrast of the images through normalization can make the SSD measure perform much better. \\

\textbf{Regularizer}: When only focusing on the distance measure, the image registration program can come up with some unrealistic scenarios (for further discussion, see \cite{Fischer_2008}). Sometimes it compresses the template image to fit within some small region of the reference. Other times it may create visible folds in the image. One way to counteract this is by introducing a regularizer.

Two commonly used regularizers implemented in FAIR are elastic and hyperelastic \cite{BurgerEtAl2013}. Both of these support the same idea but vary in severity. An elastic regularizer will try to keep the image registration close to the original orientation by introducing linear strain to any change. A hyperelastic regularizer does this with a nonlinear strain equation.
Because of this, the elastic regularizer is only accurate for smaller displacements. For anything larger, the elastic regularizer risks becoming inaccurate and causing folding in the transformation grid. The more realistic, yet computationally more expensive hyperelastic regularizer also prevents folding. Either of these regularizers would work for the Chiari data, but hyperelastic was used since seems to remove more of the edge cases and its computational costs for two-dimensional problems is feasible. \\

\textbf{Regularization Parameter}: When constructing the objective functional from \textbf{Equation \ref{joint_objective}}, there is a trade-off between whether to minimize the distance or follow the regularizer. The parameter $\alpha$ controls the balance between allowing for displacement between the original and transformed image, to better the distance measure, or to penalizing displacement, to preserve the general shape of the image throughout the registration \cite{Modersitzki2009}. 
Note that in the objective functional equation, $\alpha$ corresponds to the regularization parameter. This coefficient was chosen to be 500 in our final algorithm based on trial and error, but the parameter selection could be further refined for better results. \\

\textbf{Multilevel Optimization}: As common in FAIR, Gauss Newton optimization was applied to a discretization of the objective functional in Equation~\ref{joint_objective} to generate a transformation from template to reference. Because discretizing objective functional on a 256x256 grid can lead to many local minima, and the convergence of this Gauss Newton to an appropriate minimum is dependent on an initial guess, a multilevel approach was used. This involves creating coarser discretizations of the problem and leverages the continuous image model that we can find through linear interpolation of the image data as detailed below.

We model the template and reference images as continuous functions, $\calT : \Omega \to \R$ and $\calR : \Omega \to \R$ where $\Omega \subseteq \R^2$ corresponds to coordinates of pixels in the image grid. Coarser representations are created by changing the grid for which our interpolated image function is called, that contain fewer pixels: in our case, three additional representations, with 32x32, 64x64, and 128x128 pixels each. The optimization is applied in a way that moves up through these representations, applying optimization on the coarsest first and using the found transformation as the initial guess for the image at the next level. 

The coarsest representation, described above, is also used in an initial pre-registration step where a parametric transformation is estimated to better align the template and reference images.  This step, much like the step to choose appropriate template images, helps simplify the registration. One natural instinct for this rough transformation may be to simply translate and rotate the template image to align the reference. This would be a rigid transformation. There are advantages to keeping the transformations rigid, especially in cases where we expect little difference in the shapes of two images, but more flexibility can be added by using an affine transformation. This allows for additional linear transformations, such as shearing and scaling. We use affine transformations because of this flexibility to further align our images of often widely varied brain shapes.

\subsection{Examples}

This subsection compares two examples of how the registration performed --- one showing a successful registration and another showing how it can fail.

\begin{figure}[t]
    \centering
    \begin{tabular}{ccc}
        & \textbf{$\calR$ with $\calM_\calT$ and $\calM_\calR$} & \textbf{$\calR$ with $\calM_\calT[y]$ and $\calM_\calR$} \\
        \vspace{3mm}
        \rotatebox[origin=c]{90}{Patient 1} &
        \raisebox{-0.5\height}{\includegraphics[width=0.35\textwidth]{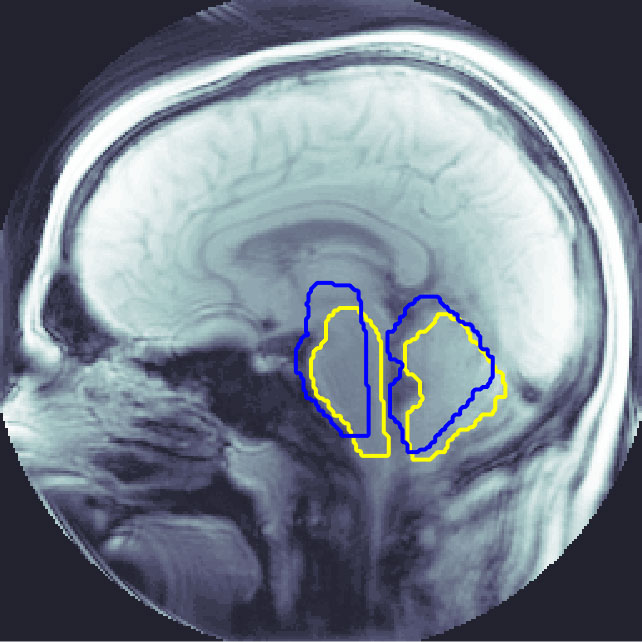}} & \raisebox{-0.5\height}{\includegraphics[width=0.35\textwidth]{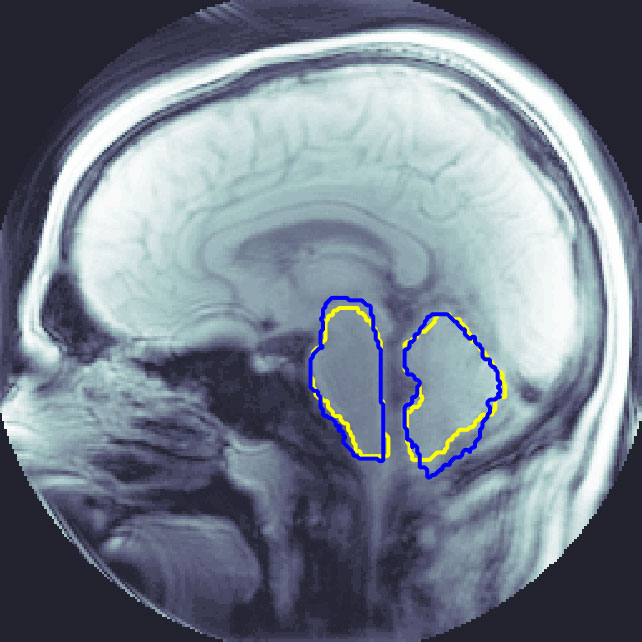}} \\
        \rotatebox[origin=c]{90}{Patient 2} &
        \raisebox{-0.5\height}{\includegraphics[width=0.35\textwidth]{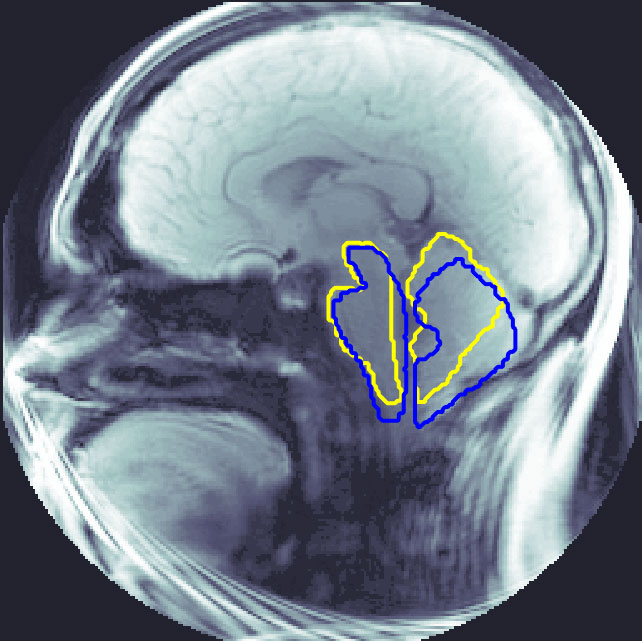}} & \raisebox{-0.5\height}{\includegraphics[width=0.35\textwidth]{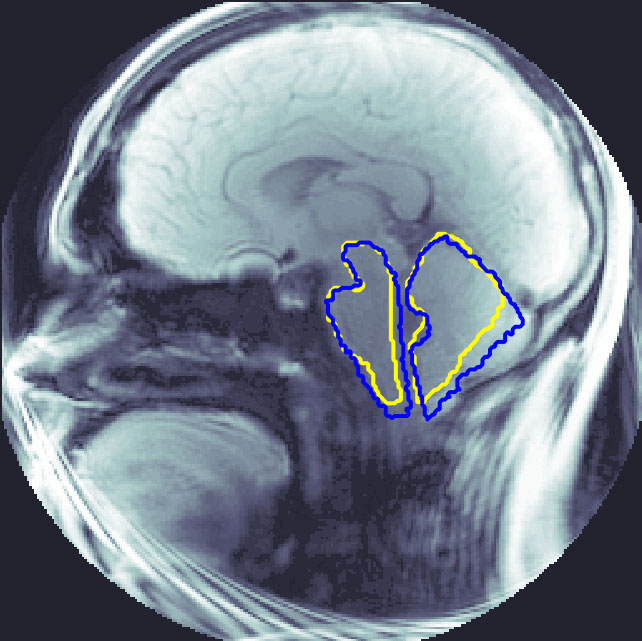}}
    \end{tabular}
    \caption{The mask transformations from patients 1 and 2. The first column of images are the reference and reference mask (yellow) with the template mask on top (blue); the second column of images are the same except with the blue outline representing the transformed template mask. The Dice similarities are in \textbf{Table \ref{patient_dice}}.}
    \label{patient_mask}
\end{figure}

Using the same parameters described in \textbf{Section \ref{Registration}}, a single registration was run on both patients. \textbf{Figure \ref{patient_mask}} compares how well the masks were aligned before and after the registration. The transformations found for each patient were reasonable and proved to be somewhat successful. However, there were still some regions that required more attention. Particularly, the lower right of the cerebellum overextends in both patients.

\begin{figure}[t]
    \centering
    \begin{tabular}{ccccc}
        &\textbf{Single $\calT[y]$} & \textbf{Average $\calT[y]$} & \textbf{Average $\calT[y] > 0.5$} & \\
        \vspace{3mm}
        \rotatebox[origin=c]{90}{Patient 1} &
        \raisebox{-0.5\height}{\includegraphics[width=0.25\textwidth]{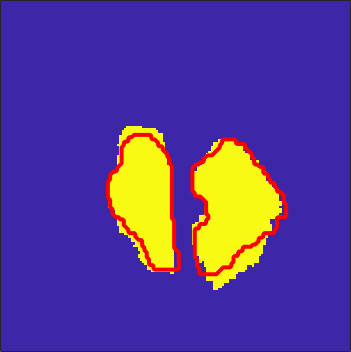}} & \raisebox{-0.5\height}{\includegraphics[width=0.25\textwidth]{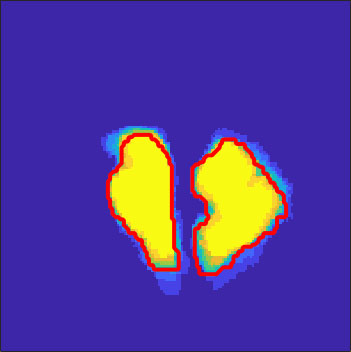}} & \raisebox{-0.5\height}{\includegraphics[width=0.25\textwidth]{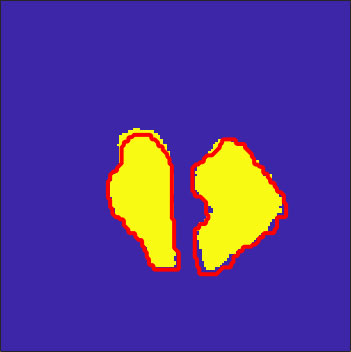}} & \multirow{2}{*}{\includegraphics[width=0.045\textwidth]{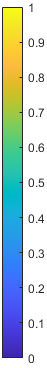}} \\
        \rotatebox[origin=c]{90}{Patient 2} &
        \raisebox{-0.5\height}{\includegraphics[width=0.25\textwidth]{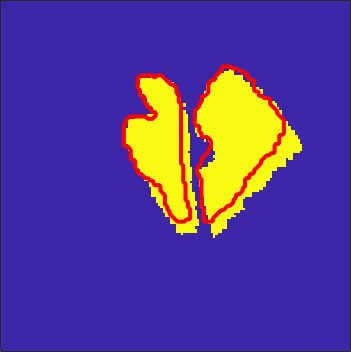}} & \raisebox{-0.5\height}{\includegraphics[width=0.25\textwidth]{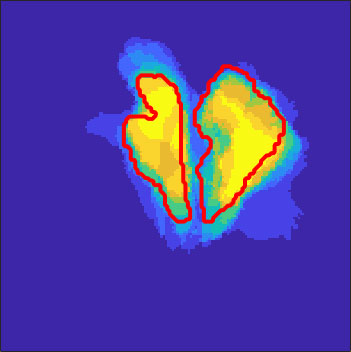}} & \raisebox{-0.5\height}{\includegraphics[width=0.25\textwidth]{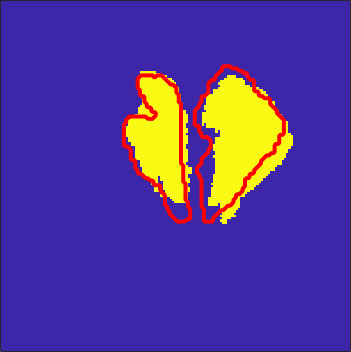}}
    \end{tabular}
    \caption{Single registration compared to the average registration program on both patients. The left column of images shows the single registration from before, the middle column is the average of all the pixels with $n = 10$ patients, and the right column is the same figure after using a threshold of $0.5$. This is the same mask as \textbf{Figure \ref{patient_mask}} and the similarity is shown under the average column of \textbf{Table \ref{patient_dice}}.}
    \label{patient_avg}
\end{figure}

In an attempt to smooth some of the outlying regions, the average program was run on both patients as shown in \textbf{Figure \ref{patient_avg}} and \textbf{Table \ref{patient_dice}}. Tuning the parameters of the average program (as seen in \textbf{Section \ref{AtlasResults}}) resulted in it performing better on most patients. Averaging this with other patients lessened the severity of this by eliminating the outliers that result from running the registration once.

\begin{table}[t]
    \centering
    \begin{tabular}{|c|c|c|c|c|c|c|}
        \hline
        & \multicolumn{3}{c|}{Patient 1} & \multicolumn{3}{c|}{Patient 2} \\
        \cline{2-7}
        & Original & Transformed & Average & Original & Transformed & Average \\
        \hline
        Brain stem & 0.734 & 0.896 & 0.928 & 0.675 & 0.845 & 0.805 \\ 
        \hline
        Cerebellum & 0.659 & 0.910 & 0.936 & 0.805 & 0.855 & 0.850 \\ 
        \hline
        Full Mask & 0.701 & 0.902 & 0.932 & 0.728 & 0.849 & 0.824 \\
        \hline
    \end{tabular}
        
    \caption{Dice similarities of the patient masks from \textbf{Figure \ref{patient_mask}} before and after the transformation along with the average version from \textbf{Figure \ref{patient_avg}}.}
    \label{patient_dice}
\end{table}

Looking at the particular patients from \textbf{Table \ref{patient_dice}}, the averaging improved the similarity of patient 1 by around 2-3\% but decreased the similarity of patient 2 by around the same amount (particularly in the brain stem). This could represent the need for a more definitive metric or a larger dataset; either of these solutions would aim to increase the quality of the chosen template.

\section{Neural-Network-Based Semantic Segmentation}

The overarching goal of this approach is to find a relationship between the input DENSE imaging data and the known target segmentation masks through a function which directly labels the image. We can define our input $I \in \R^{256\times256\times3}$ and masks $M \in \{0,1,2\}^{256 \times 256}$. As stated previously in \textbf{Section \ref{Notation}}, $M$ symbolizes the known target masks while $M_{p}$ corresponds to the model predicted mask. 

\subsection{U-Net: A Convolutional Neural Network}

Given an input image $I$, our goal is to estimate a three-dimensional tensor $P$ in the set
\begin{equation}
    \Delta = \left\{Q \in \R^{256 \times 256 \times 3} : Q_{i,j,k} \geq 0 \text{ }\forall \text{ } i,j,k \text{ and } \sum_{k=0}^{2} Q_{i,j,k}=1\right\},
\end{equation}
where $P_{i,j,k}$ is the probability the pixel at $i,j$ belongs to class $k$. The relationship between $I$ and $P$ can then be represented by a parameterized function
\begin{equation}
    f_{\theta}: \R^{256\times 256\times 3} \to \Delta,
\end{equation}
where $\theta$ represents the weights of the model. 
To then create a mask $M_p$, each pixel is assigned to the class with the highest probability; i.e., we maximize over the third dimension in $P$.

\begin{figure}[t]
    \centering
    \includegraphics[width=.9\textwidth]{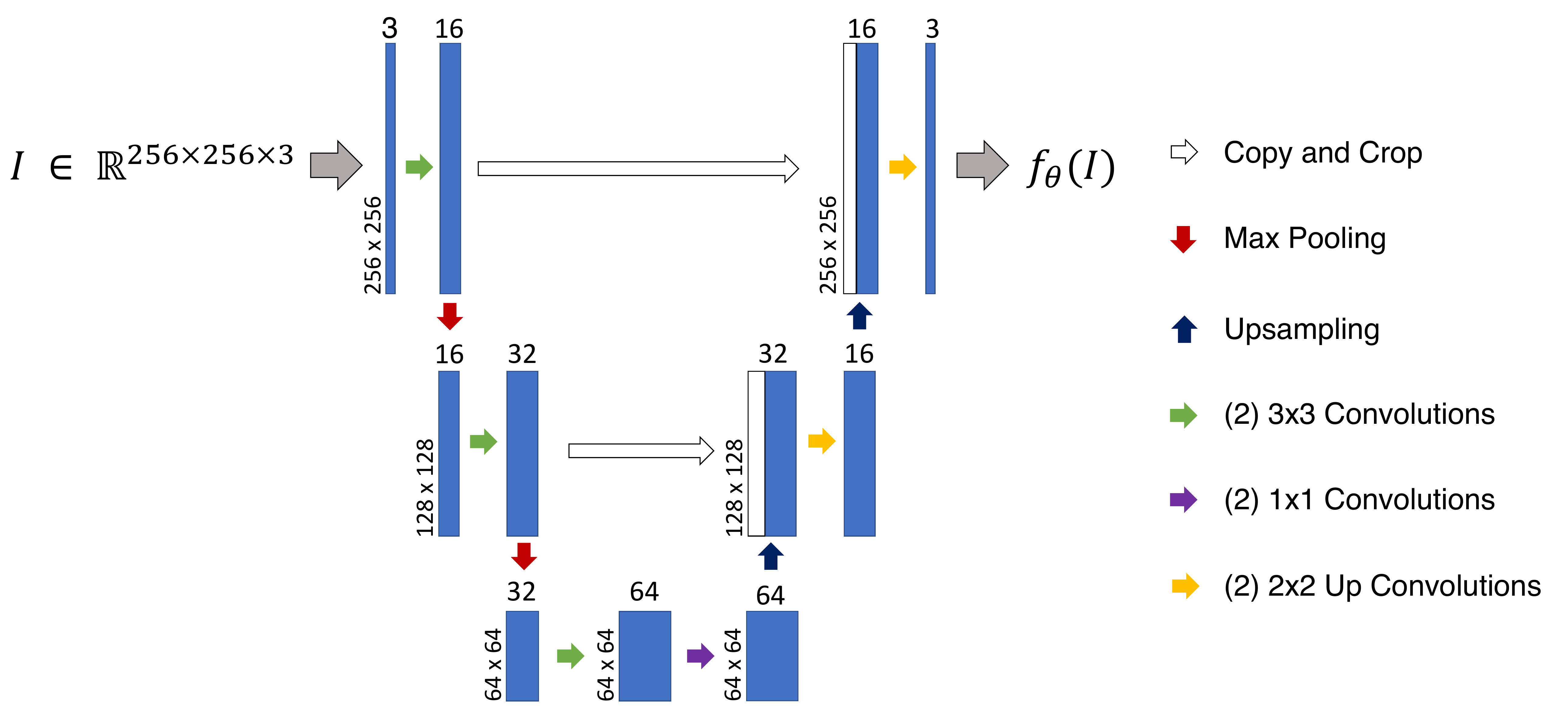}
    \caption{U-Net architecture used in our study based on the figure provided in U-Net: Convolutional Networks for Biomedical Image Segmentation \cite{RonnebergerEtAl2015}. Here numbers placed on top of the boxes represent number of channels while those on the side represent image dimensions.}
    \label{fig:Unet-Arch}
\end{figure}

Our model is a convolutional neural network (CNN). 
CNNs are a type of neural network which process data that have a ``grid like topology," such as image data (a 2D grid of pixels), and utilizes a convolution operator. Since the data in this study consist of multidimensional arrays of pixels and the goal is to correctly predict which class each pixel belongs to, a CNN is a natural choice. Many deep networks require thousands of images, however, our data set, as is common in biomedical applications, is much smaller. 
A CNN architecture that has been successful in other biomedical segmentation tasks with limited amounts of training data is the U-Net~\cite{RonnebergerEtAl2015}. A U-Net consists of two main sections: 
the contracting path (encoder) and the expansive path (decoder) shown on  the left and right side of \textbf{Figure \ref{fig:Unet-Arch}}, respectively.

In the first layer of the contracting path, a 3x3 convolution is applied to each channel increasing the number of total feature channels. This is followed by a max pooling operation to break down the image resolution by decreasing image size to better identify features. A similar convolution structure is applied at each layer doubling the number feature channels. This leads to the bottom most layer where 1x1 convolutions are used, maintaining number of channels. Increasing the size of the image data, the expansive path (decoder) replaces the max pooling with upsampling operators. In the same formation as the contracting path, 2x2 up convolutions are applied except these halve the number of feature channels. During the process, high resolution features found during encoding are `copied and cropped' and combined with the decoded output increasing dimension size and providing information that will be used in the convolution layer to produce more accurate outputs.

As the input images were run through the U-Net, the network used information from each channel to assign each pixel a probability for each class. These values are used to calculate the loss and to find the final predicted segmentation, $M_p$. 

\subsection{Training the Model}

In order to get a successful model, we want the predicted mask $M_p$ and the target mask $M$ to be as similar as possible. The first step in maximizing similarities is to create a satisfactory model. This was done through supervised training of the network using a training set, $\{\left(I_{t},M_{t}\right)\}_{t=1,2,\ldots41}$, a subset of the 51 DENSE MR images and corresponding masks $\{(I,M)\}$. To test the accuracy of the model on `new' data, a validation set, $\{\left(I_{v},M_{v}\right)\}_{v=1,2,\ldots,10}$, was defined using the remaining 10 DENSE images and masks not included in training. 

The U-Net was setup based on code provided by Aman Arora \cite{UNetCode}. The network was then trained by looping a sequence of inputting $I_{t}$ to the U-Net, computing the loss between $f_{\theta}(I_{t})$ and $M_{t}$, and updating the model with the optimizer for a set number of iterations. In each iteration, the updated model was also run on the validation data to gauge its generalization; the validation loss was used to tune hyperparameters but was neither used to calculate gradients nor to change model parameters.

Comparing the probability matrix $P = f_{\theta}(I)$ to the corresponding $M$, for both training and validation data, shows the success of the model and can be numerically calculated using a loss function.

By maximizing the probability at the class corresponding to that of the known mask in the same position, the model error is minimized. In other words, our goal is to maximize
\begin{equation}\label{eq:lossMax}
    [f_{\theta}(I)]_{i,j,[M]_{i,j}}, \quad \text{for all} \quad i = 1, \ldots, 256, j=1,\ldots,256,
\end{equation}
which is equivalent to minimizing 
\begin{equation}\label{eq:lossMin}
     -\log \left( [f_{\theta}(I)]_{i,j,[M]_{i,j}} \right), \quad \text{for all} \quad i,j \in [1,256],
\end{equation}
where $i,j$ corresponds to the pixel position.

Averaging \textbf{Equation \ref{eq:lossMin}} over all training image sand all pixels results in the objective function
\begin{equation}\label{eq:objFunct}
    {\cal{J}}_{CNN}(\theta) = \frac{1}{41\cdot256^{2}} \sum_{t=1}^{41}\sum_{i,j = 1}^{256} -\log\left([f_{\theta}(I_{e})]_{i,j,[M_{e}]_{i,j}}\right).
\end{equation}

It can be seen that ${\cal{J}}_{CNN}$ equals the cross-entropy loss between $[f_{\theta}(I)]_{i,j,:}$ and the probability distribution defined by the standard basis vector, $[M]_{i,j}$, which calculates how far the model prediction is from the expected output.

Inputting $f_{\theta}(I)$ and $M$ in the loss function is a forward pass; this is done for both training and validation data. By executing a backwards pass on the loss, gradients can be calculated and used by the optimizer to update model parameters. As emphasized before, a backward pass was only executed with respect to the training data.

A limited-memory Broyden–Fletcher–Goldfarb–Shanno (lBFGS) algorithm from the PyTorch library \cite{paszke2017automatic} was used to optimize the overall loss function, shown previously in \textbf{Equation \ref{eq:objFunct}}. lBFGS is a Quasi-Newton method which uses a line search to automatically find the optimal learning rate $\epsilon$ \cite{Goodfellow:2016wc} which satisfies the strong Wolfe conditions set in the algorithm parameters. The linesearch ensures decrease of the  objective function while keeping $\epsilon$ from becoming too small.

The lBFGS algorithm uses the gradient calculated from the loss function and the learning rate found in the line search to update the model weights. The images were then, once again, run through the U-Net in hopes of finding a better segmentation. The model with the lowest validation loss was saved. 

\section{Results}
This section begins with results from the processes of creating the atlas-based and neural-network-based methods; overall results on the final testing set follow.

\subsection{Atlas-Based Image Registration Results}
\label{AtlasResults}

When creating the averaging program, two main parameters were chosen arbitrarily: $n$ and threshold. The $n$ refers to the number of template images to average and the threshold refers to which values should be counted when converting the average image to binary. In order to decide on a more definitive value for these parameters, a program was made to iterate through different possible values and find which performed the best.

\begin{figure}[t]
    \centering
    \begin{tabular}{cc}
        \multirow{2}{*}{\textbf{Average Dice vs $n$}} & \textbf{Average Absolute Difference} \\
        & \textbf{in Biomarker vs $n$} \\
        \hspace{-5mm}
        \includegraphics[width=0.48\textwidth]{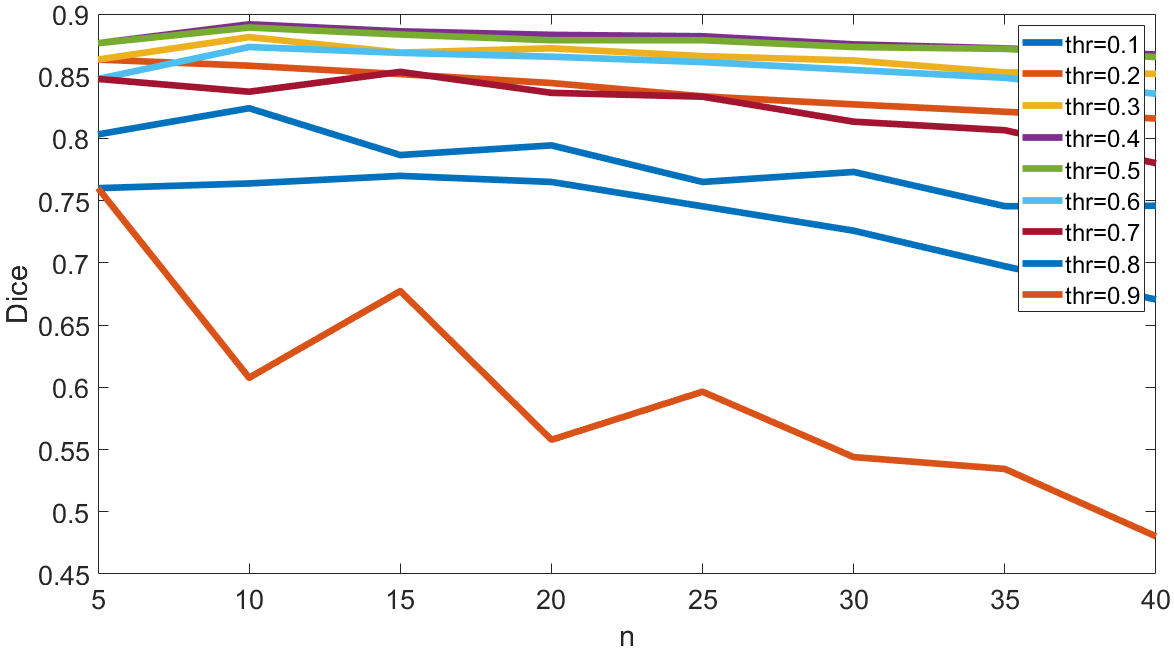} & \includegraphics[width=0.48\textwidth]{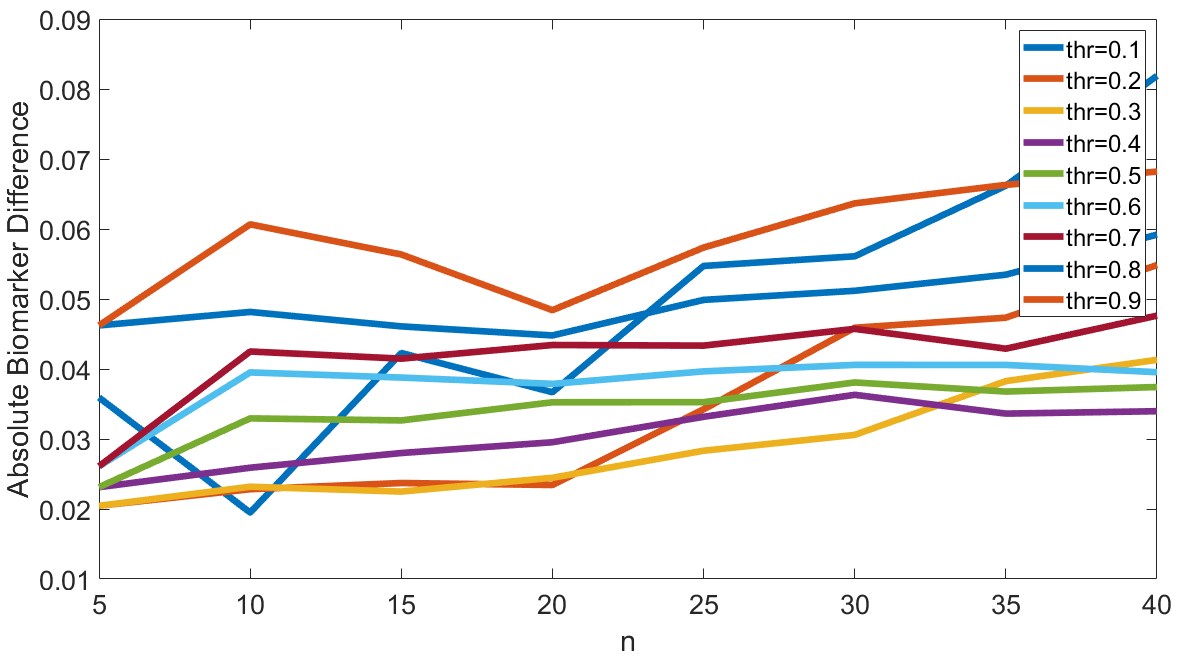}
    \end{tabular}
    \caption{The left image shows the average Dice index plotted against the number of templates used in registration, and the right shows the average absolute difference between predicted and actual biomarker plotted against the number of templates used in registration. For both these figures, the average was computed over all 41 brains in the training set. The different curves correspond to different thresholds chosen for registration.}
    \label{choosing_n}
\end{figure}

This program was run over two metrics for how well it performed: Dice similarity and biomarker difference. Both of these metrics compare the atlas-based generated mask (average program) to the manually segmented mask and then average together all of the patients. These produced approximately the same results since the biomarker difference depends on how similar the masks are. A plot showing averaged Dice indices and absolute biomarker error over changes of both $n$ and the threshold can be seen in \textbf{Figure \ref{choosing_n}}.

To choose these $n$ and threshold parameters, we performed a grid search, first using a wide range of $n$ and threshold values (ranging from 5 to 40 and 10\% to 90\%, respectively). We chose to complete all further registrations with $n=10$ and a threshold of 50\%, as this maximizes the average Dice index. The maximizer of the Dice index was chosen over the minimizer of the biomarker error because of how sensitive this biomarker measure is to slightly different segmentations. We expect that the behavior of the Dice index average will generalize better to other data.

\subsection{Neural-Network-Based Semantic Segmentation Results}

\begin{figure}[t]
    \centering
    \includegraphics[width=1\linewidth]{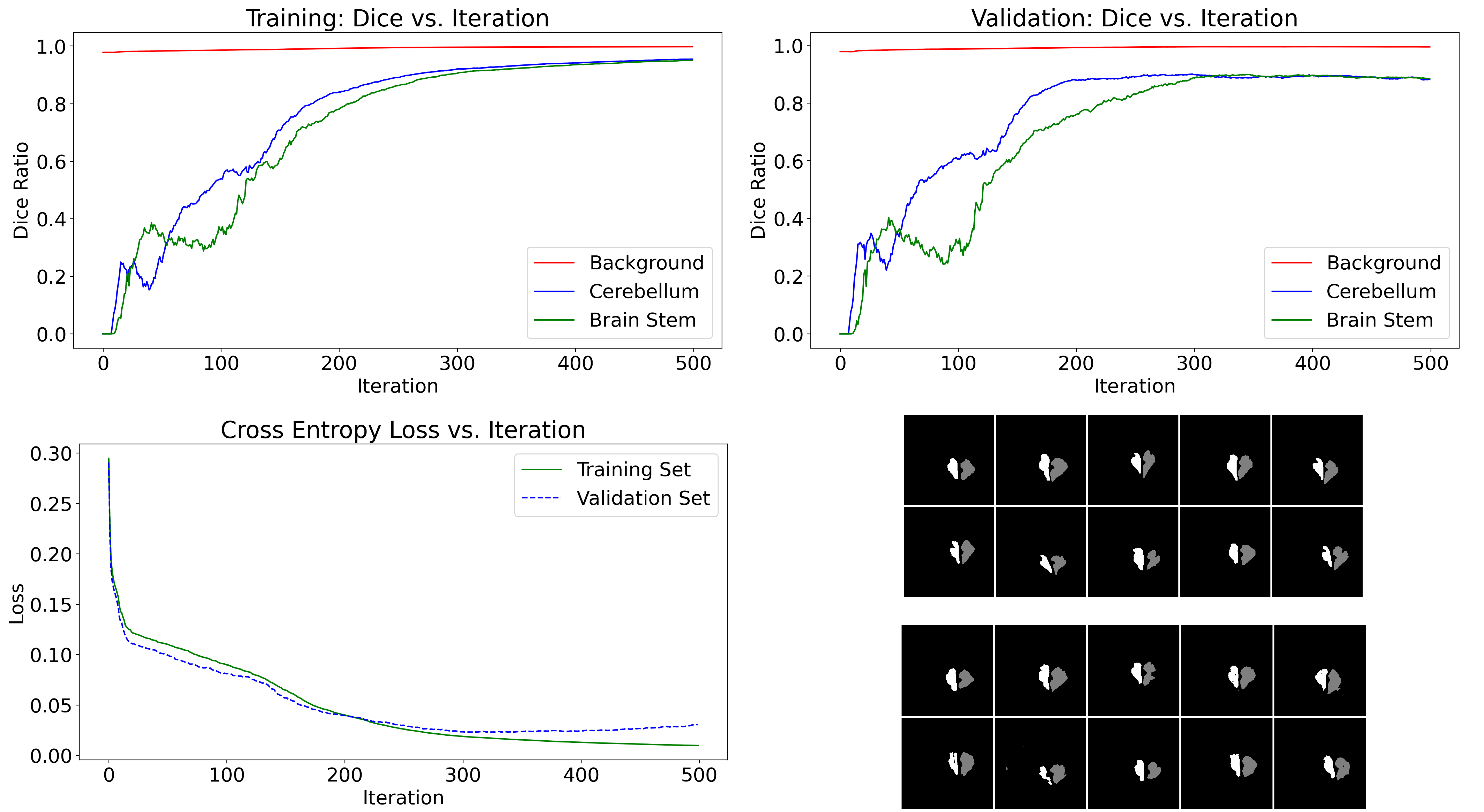}
    \caption{Dice similarities of each class for training (top left) and validation (top right) data over training iterations, cross-entropy loss for each set over training iterations (bottom left), and a comparison of true (top) to predicted (bottom) masks for validation data (bottom right).}
    \label{fig:MLPlots}
\end{figure}

While training the model, cross-entropy loss and Dice similarities were used as a measure of model success. Our final model was saved at the lowest validation loss found, which occurred at iteration 309, and was used to produce the final results shown below. Cross-entropy loss and Dice similarities throughout model training are plotted in \textbf{Figure \ref{fig:MLPlots}} alongside a comparison of true to predicted masks for validation data using the final model. Final results of the model used can be found in \textbf{Table \ref{tab:mlComp}}.

\begin{table}[t]
    \centering
    \begin{tabular}{|c|c|c|c|c|}
        \hline    
         & Cross-Entropy & \multicolumn{3}{c|}{Dice} \\
         \cline{3-5}
         & Loss & Background & Cerebellum & Brain Stem\\
         \hline
         Training Data & 0.0180 & 0.9962 & 0.9216 & 0.9116 \\
         \hline
         Validation Data & 0.0229 & 0.9953 & 0.8965 & 0.8919\\
         \hline
    \end{tabular}
    \caption{Cross-Entropy loss and Dice similarity of each class for training and validation data sets for the neural-network based segmentation. The results shown here correspond to the loss and the respective Dice value in \textbf{Figure \ref{fig:MLPlots}} at iteration 309 where minimum validation loss was found.}
    \label{tab:mlComp}
\end{table}

The model was then used on the test data previously set aside. Biomarkers were calculated using test data segmentations which were compared to the corresponding biomarkers produced in the atlas-based method and those from the known masks $M$.
  
 \subsection{Comparison Results}  

\begin{figure}[t]
    \centering
    \includegraphics[width=1\textwidth]{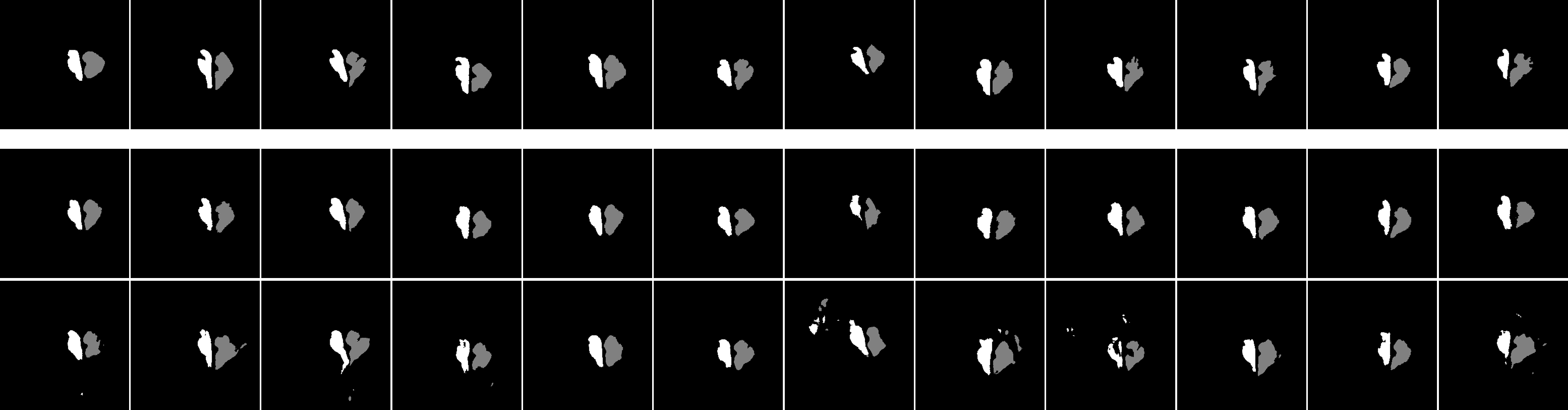}
    \caption{Comparison of the true masks (top), atlas-based predicted (middle), and neural-network-based predicted (bottom) for all subjects in the test data.}
    \label{fig:result-masks}
\end{figure}

When analyzing atlas-based (AB) and neural-network-based (NN) in the end, we relied on the Dice and biomarker (BM) metrics. \textbf{Table \ref{tab:beast}} shows the samples provided in the testing set and compares these methods to the manually segmented mask (true). The masks that were compared are shown in \textbf{Figure \ref{fig:result-masks}}.

\begin{table}[t]
    \centering
    \begin{tabular}{|c|c|c|c|c|c|c|c|c|c|c|}
        \hline
        Test Set &
         \multicolumn{4}{c|}{Dice Similarity Index}&
         \multicolumn{6}{c|}{Displacement Biomarker}
         \\
         \cline{2-11}
        Subject &
         \multicolumn{2}{c|}{Brain Stem}&
         \multicolumn{2}{c|}{Cerebellum} &
         \multicolumn{3}{c|}{Brain Stem} & 
         \multicolumn{3}{c|}{Cerebellum} \\
         \cline{2-11}
         \# & AB & NN & AB & NN & True & AB & NN & True & AB & NN \\
         \hline
         1 & 0.70 & \textbf{0.83} & 0.81 & \textbf{0.84} & 2.32 & \textbf{2.37} & 2.25 & 1.97 & 1.31 & \textbf{1.62} \\
         \rowcolor{Gray}
         2 & 0.87 & \textbf{0.90} & \textbf{0.89} & 0.86  & 0.69 & 0.68 & \textbf{0.68} & 0.47 & 0.45 & \textbf{0.45} \\
         3 & 0.81 & \textbf{0.84} & 0.87 & \textbf{0.88} & 1.25 & 1.13 & \textbf{1.15} & 1.34 & 1.03 & \textbf{1.22} \\
         \rowcolor{Gray}
         4 & \textbf{0.84} & 0.81 & \textbf{0.88} & 0.71 & 1.06 & 0.99 & \textbf{1.00} & 0.58 & 0.53 & \textbf{0.54} \\
         5 & 0.82 & \textbf{0.88} & 0.88 & \textbf{0.90} & 1.02 & \textbf{0.99} & 0.98 & 0.64 & 0.58 & \textbf{0.59} \\
         \rowcolor{Gray}
         6 & 0.53 & \textbf{0.74} & 0.60 & \textbf{0.67} & 0.85 & \textbf{0.77} & 0.76 & 0.54 & 0.90 & \textbf{0.60} \\
         7 & 0.90 & \textbf{0.94} & 0.92 & \textbf{0.93} & 0.86 & 0.84 & \textbf{0.85} & 0.48 & 0.45 & \textbf{0.47} \\
         \rowcolor{Gray}
         8 & 0.87 & \textbf{0.90} & 0.90 & \textbf{0.96} & 0.89 & \textbf{0.89} & 0.88 & 0.48 & 0.51 & \textbf{0.49} \\
         9 & \textbf{0.91} & 0.90 & \textbf{0.90} & 0.87 & 0.73 & \textbf{0.73} & 0.74 & 0.52 & \textbf{0.51} & 0.51 \\
         \rowcolor{Gray}
         10 & 0.83 & \textbf{0.85} & \textbf{0.92} & 0.87 & 1.05 & \textbf{0.97} & 1.20 & 1.24 & \textbf{1.07} & 0.97 \\
         11 & 0.84 & \textbf{0.88} & 0.85 & \textbf{0.91} & 1.48 & \textbf{1.48} & 1.47 & 0.99 & 0.85 & \textbf{0.90} \\
         \rowcolor{Gray}
         12 & \textbf{0.87} & 0.82 & 0.92 & \textbf{0.93} & 0.58 & \textbf{0.58} & 0.59 & 0.37 & 0.39 & \textbf{0.37} \\
    \hline
    \end{tabular}
    \caption{Results of atlas-based (AB) and neural-network-based (NN) semantic segmentation approaches on the testing set. The bold-face numbers represent the best results for each subject (closest to $1$ for Dice similarity, closest to true value for the biomarkers).}
    \label{tab:beast}
\end{table}

\begin{figure}[t]
    \centering
    \begin{tabular}{cc}
        \multicolumn{2}{c}{\textbf{Dice similarity to true masks}}\\
        {}&{}\\
        Brain Stem & Cerebellum\\
        \includegraphics[width=.5\textwidth]{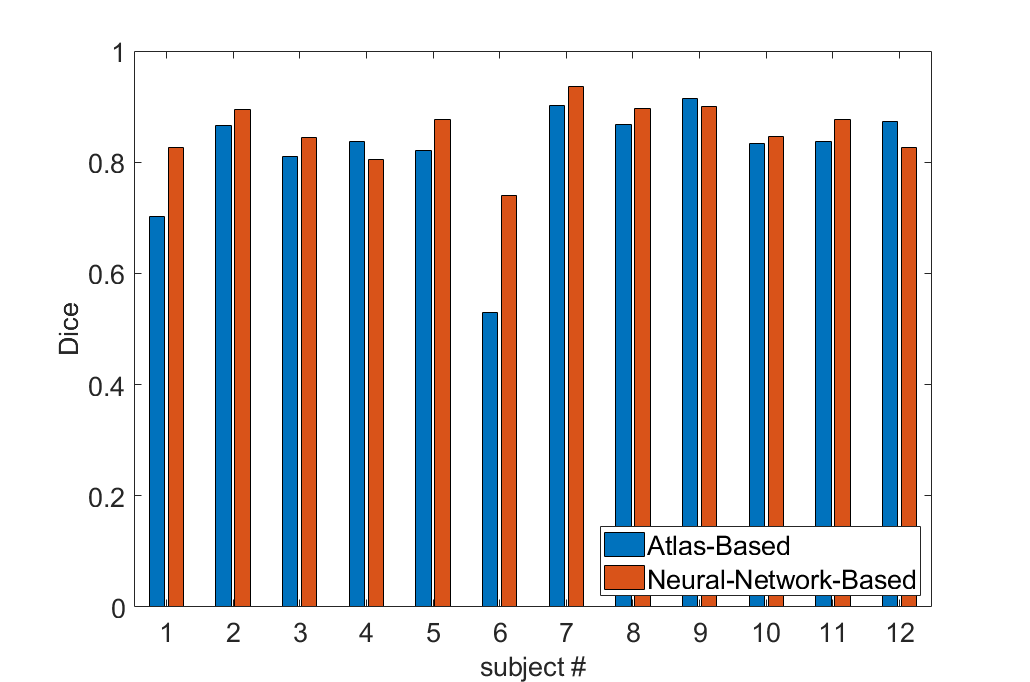} & 
        \includegraphics[width=.5\textwidth]{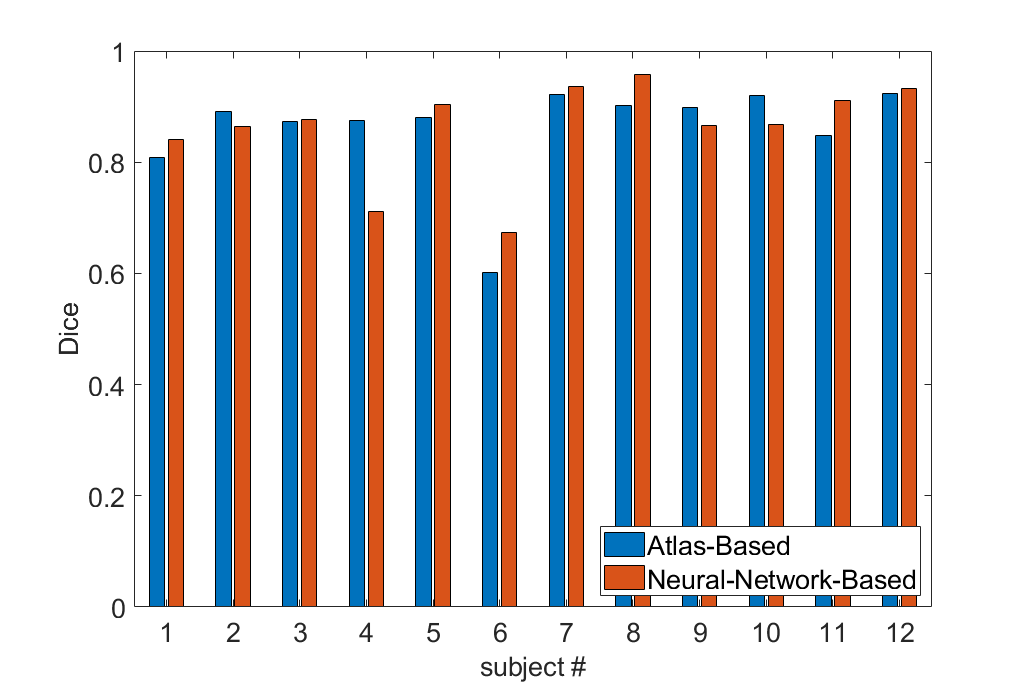}\\
        {}&{}\\
         \multicolumn{2}{c}{\textbf{Relative biomarker difference from true values}}\\
         {}&{}\\
         Brain Stem & Cerebellum\\
         \includegraphics[width=.5\textwidth]{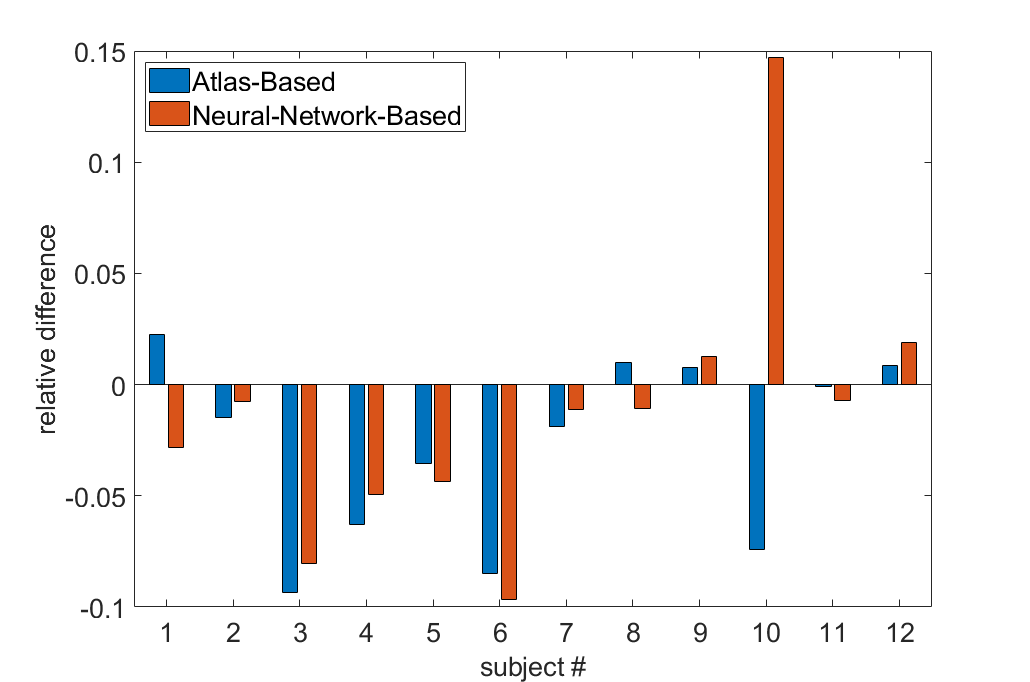} & 
        \includegraphics[width=.5\textwidth]{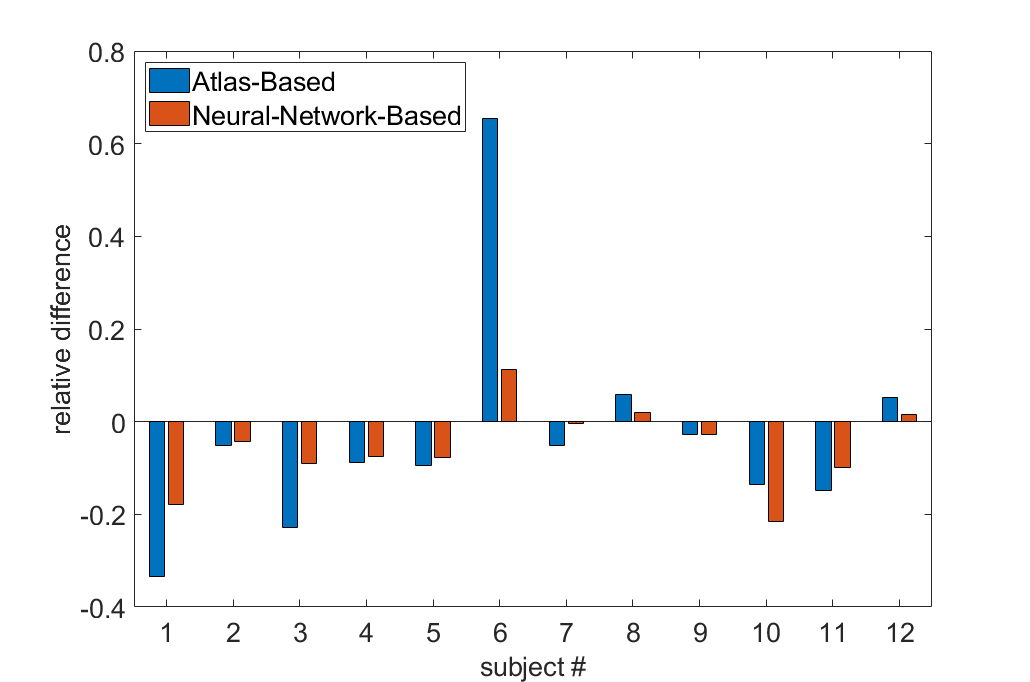}\\
        
    \end{tabular}
    \caption{Bar graph displaying atlas-based (blue) and neural-network-based (orange) results from testing set in terms of Dice similarity and relative biomarker error when compared to target (manually generated) results. Note that higher values are better for the Dice similarity while a lower difference is desirable for biomarkers. Negative relative differences indicate that manually generated results are larger than those predicted (and vice versa).}
    \label{fig:test-result-plot}
\end{figure}

\begin{figure}[t]
    \centering
    \begin{tabular}{c}
        \textbf{Relative Biomarker Difference versus Dice Similarity}\\
        \includegraphics[width=.8\textwidth]{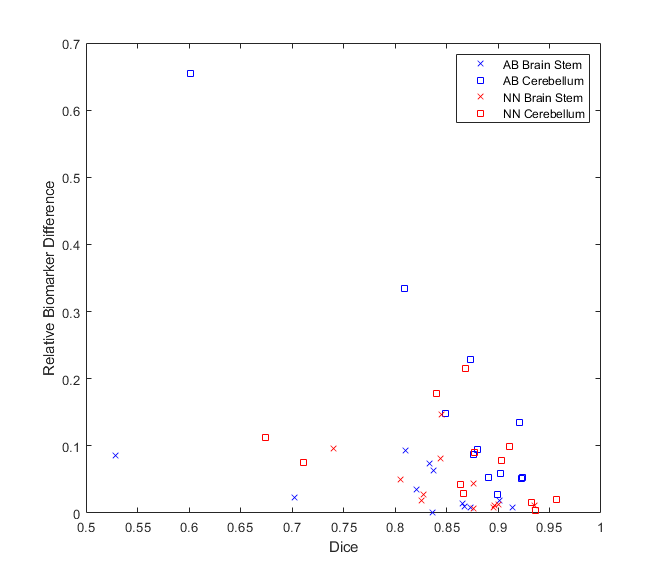}\\
    \end{tabular}
    \caption{Scatter plot showing Dice similarity versus absolute relative biomarker difference split between both regions (x's for brain stem, squares for cerebellum) and both methods (blue for atlas-based, red for neural-network-based).}
    \label{fig:full-scatter}
\end{figure}

\begin{figure}[t]
    \centering
    \begin{tabular}{c}
        \textbf{Error Spread in Dice and Biomarker Similarities}\\
        \includegraphics[width=.8\textwidth]{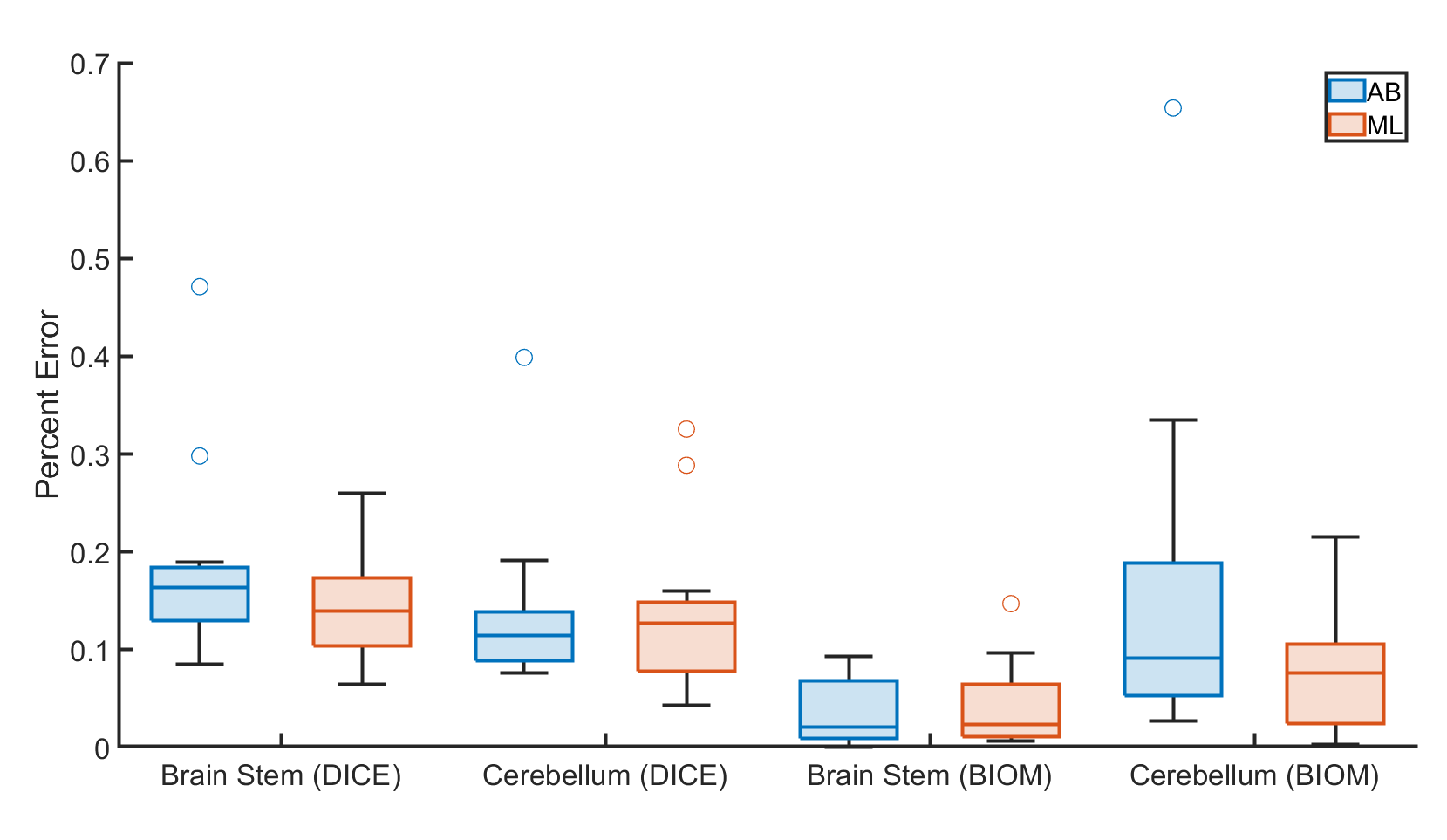}\\
    \end{tabular}
    \caption{Box plot showing errors of predicted masks and biomarkers. The Dice error (left) is calculated as $1 - D_{Dice}$ between the manually and automatically produced masks. The biomarker error (right) is a relative error between the manually and automatically produced biomarkers ($|\text{prediction} - \text{true}| / \text{true}$).  These errors are split between the brain stem and cerebellum, where blue indicates the errors of the atlas-based method and red indicates the error of the neural-network-based method.}
    \label{fig:box}
\end{figure}

The bars chart in \textbf{Figure \ref{fig:test-result-plot}} can offer more visual insight into which method did better on each patient. Some other representations and related results of \textbf{Table \ref{tab:beast}} are shown in \textbf{Table \ref{tab:analytics}}, \textbf{Figure \ref{fig:full-scatter}} and \textbf{Figure \ref{fig:box}}.

\begin{table}[t]
    \centering
    \begin{tabular}{|l|c|c|c|c|c|c|c|c|c|c|}
        \hline
         \multirow{2}{*}{Analytics} &
         \multicolumn{2}{c|}{BS DICE} &
         \multicolumn{2}{c|}{CB DICE} &
         \multicolumn{3}{c|}{BS BIOM} & 
         \multicolumn{3}{c|}{CB BIOM} \\
         \cline{2-11}
          & AB & NN & AB & NN & True & AB & NN & True & AB & NN \\
         \hline
         Mean: & 0.82 & \textbf{0.86} & \textbf{0.86} & 0.86 & 1.06 & 1.04 & \textbf{1.05} & 0.80 & 0.71 & \textbf{0.72} \\
         Avg. Error: & 0.18 & \textbf{0.14} & \textbf{0.14} & 0.14 & --- & \textbf{0.04} & 0.04 & --- & 0.16 & \textbf{0.04} \\

    \hline
    \end{tabular}
    \caption{Summary of results on testing set. Dice error computed as $1-D_{Dice}$. Biomarker error (relative) computed as $|\text{prediction} - \text{true}| / \text{true}$.}
    \label{tab:analytics}
\end{table}

\section{Discussion}
In this paper, we developed and compared two methods -- one atlas-based and one neural-network-based -- that identify the cerebellum and brain stem in a given MR image.
The goal of the segmentation is to produce a displacement biomarkers that can aid in the diagnosis of CMI. After developing and evaluating these methods, we found interesting differences between the accuracy of each method in terms of the mask similarity and biomarker similarity. 

In 9 of the 12 test cases for the brain stem and 8 out of 12 cases for the cerebellum, the neural-network-based method produced higher similarity masks (\textbf{Table \ref{tab:beast}}). For the brain stem, this corresponded to an average mask similarity that was higher for the neural-network-based method. However, the atlas-based method produced slightly more similar masks \textit{on average} for the cerebellum than the neural-network-based method, even though it performed relatively worse in a majority of cases. 

In terms of biomarker similarity, although the neural-network-based approach produced better segmentations (in 9 of 12 cases and on average), the atlas-based method produced more accurate biomarkers in 8 of 12 cases and on average in terms of lowest relative error. For the cerebellum, the neural-network-based method produced better biomarkers in 10 of 12 cases, and on average.

These differences between the Dice similarity and biomarker similarity for the brain stem show that although the neural-network produced segmentations that were more similar to those manually drawn in a pixel-wise sense, the atlas-based method's segmentations led to an average of displacement values that more closely matched the average displacement biomarker from manual segmentations. The differences between which method performed best in a majority of cases versus on average are due to the spreads of error across these test subjects. \textbf{Figures \ref{fig:test-result-plot}} and \textbf{\ref{fig:box}} and \textbf{Tables \ref{tab:beast}} and \textbf{\ref{tab:analytics}} present these results in further detail.

The methods produce comparable results, and if a just single method were to be chosen to complete all analyses, we would recommend the neural-network-based approach, as it produces the best segmentations and biomarkers on average, notably outperforming the atlas-based method overall. However, a combination of the two -- with the atlas-based method used to produce the brain stem biomarker and the neural-network-based method used to produce the cerebellum biomarker -- may be more likely produce the best displacement biomarkers for a new patient based on the above results.

We anticipate that further work and further data may improve the accuracy of the neural-network-based segmentation more quickly than it would atlas-based. This further work on the neural-network-based method could include adding a regularizer, similar to that used in the atlas-method (described in \textbf{Section \ref{Registration}}), that could eliminate non-contiguous pixels from predicted masks, resulting in even better segmentations. A multi-batch approach, that iteratively creates training and validation sets across non-testing data, could also further help improve results, to maximally take advantage of the currently relatively small data set.

As more DENSE images are collected and labeled, both methods may produce better results; the neural-network can be trained with a larger data set, further increasing the accuracy of the segmentations, and the bank of images the atlas-based method compares a new reference to could be more likely to find a close match in a larger bank of images that could also improve the registration and resulting masks.

An interesting possibility for further work may also lie in examining the accuracy of radiologist predicted masks. Though they were treated as truth in this study, manually drawn segmentations can vary in perhaps significant ways. In the context of Chiari diagnosis with a displacement biomarker, to draw accurate masks of the brain stem and cerebellum on a new MRI is an especially challenging problem; both the cerebellum and brain stem are bordered by regions of high movement cerebrospinal fluid (CSF). If borders are drawn too wide, those high-movement pixels may be included in the displacement averages and may inflate them inaccurately. We experimented with calculating biomarkers while algorithmically ignoring values above a certain threshold as a way to ignore high movement cerebrospinal fluid. This method did not help us match manually predicted biomarkers, but it may be a low-cost and helpful method to implement even when an expert radiologist is drawing segmentations to make them more accurately include only areas with brain-tissue scale movements.

\section*{Acknowledgements}
We want to sincerely thank Dr. Lars Ruthotto and Justin Smith for their mentorship and help throughout the project, as well as Dr. John Oshinski for providing us with this data set and collaborating with us. We would further like to thank all the mentors at Emory's 2021 REU/RET Program. This work was supported by the US National Science Foundation award DMS 2051019.

\bibliographystyle{abbrv}

\bibliography{main}

\begin{thebibliography}{10}

\bibitem{AANS}
AANS.
\newblock Chiari malformation, July 2021.

\bibitem{UNetCode}
A.~Arora.
\newblock U-net: A pytorch implementation in 60 lines of code.
\newblock Sept 2020.

\bibitem{BologneseEtAl2019}
P.~A. Bolognese, A.~Brodbelt, A.~B. Bloom, and R.~W. Kula.
\newblock Chiari i malformation: Opinions on diagnostic trends and
  controversies from a panel of 63 international expert.
\newblock {\em World Neurosurgery}, pages e9--e16, 2019.

\bibitem{BurgerEtAl2013}
M.~Burger, J.~Modersitzki, and L.~Ruthotto.
\newblock A hyperelastic regularization energy for image registration.
\newblock {\em SIAM Journal on Scientific Computing}, 35(1):B132--B148, 2013.

\bibitem{Coste}
A.~Coste.
\newblock Image processing: Histograms.
\newblock 2012.

\bibitem{Fischer_2008}
B.~Fischer and J.~Modersitzki.
\newblock Ill-posed medicine{\textemdash}an introduction to image registration.
\newblock {\em Inverse Problems}, 24(3):034008, May 2008.

\bibitem{Goodfellow:2016wc}
I.~Goodfellow, Y.~Bengio, and A.~Courville.
\newblock {\em {Deep Learning}}.
\newblock MIT Press, Nov. 2016.

\bibitem{phasewrap}
T.~Lan, D.~Erdogmus, S.~Hayflick, and J.~Szumowski.
\newblock Phase unwrapping and background correction in \text{MRI}.
\newblock Proceedings of the 2008 IEEE Workshop on Machine Learning for Signal
  Processing, MLSP 2008, pages 239 -- 243, November 2008.

\bibitem{Modersitzki2009}
J.~Modersitzki.
\newblock {\em {FAIR: flexible algorithms for image registration}}, volume~6 of
  {\em Fundamentals of Algorithms}.
\newblock Society for Industrial and Applied Mathematics (SIAM), Philadelphia,
  PA, 2009.

\bibitem{Monteux2019}
A.~Monteux.
\newblock Metrics for semantic segmentation.
\newblock May 2019.

\bibitem{NwotchouangEtAl2020}
B.~S.~T. Nwotchouang, M.~S. Eppelheimer, S.~H. Pahlavian, J.~W. Barrow, D.~L.
  Barrow, D.~Qiu, P.~A. Allen, J.~N. Oshinski, R.~Amini, and F.~Loth.
\newblock {Regional Brain Tissue Displacement and Strain is Elevated in
  Subjects with Chiari Malformation Type I Compared to Healthy Controls: A
  Study Using DENSE MRI}.
\newblock {\em Annals of Biomedical Engineering}, pages 1--15, Dec. 2020.

\bibitem{paszke2017automatic}
A.~Paszke, S.~Gross, S.~Chintala, G.~Chanan, E.~Yang, Z.~DeVito, Z.~Lin,
  A.~Desmaison, L.~Antiga, and A.~Lerer.
\newblock Automatic differentiation in pytorch.
\newblock 2017.

\bibitem{RonnebergerEtAl2015}
O.~Ronneberger, P.~Fischer, and T.~Brox.
\newblock U-net: Convolutional networks for biomedical image segmentation.
\newblock In N.~Navab, J.~Hornegger, W.~M. Wells, and A.~F. Frangi, editors,
  {\em Medical Image Computing and Computer-Assisted Intervention -- MICCAI
  2015}, pages 234--241, Cham, 2015. Springer International Publishing.

\end{thebibliography}
\end{document}